\documentclass[useAMS,usenatbib]{mn2e} 
 \usepackage{graphicx}

 \title[The local standard of rest]
 {The local standard of rest from data on young objects with
 account for the Galactic spiral density wave}
 \author[V. V. Bobylev and A. T. Bajkova]{V. V. Bobylev$^{1,2}$
 \thanks{E-mail: vbobylev@gao.spb.ru} and A. T. Bajkova$^{1}$\\
 $^{1}$Central (Pulkovo) Astronomical Observatory of RAS, 65/1 Pulkovskoye Chaussee, Saint Petersburg, 196140, Russia\\
 $^{2}$ Sobolev Astronomical Institute, St. Petersburg State University, Bibliotechnaya pl. 2, St. Petersburg, 198504, Russia}

\begin{document}
 \date{Accepted 2014 March 18. Received 2014 March 16; in original form 2013 September 19}
 \pagerange{\pageref{firstpage}--\pageref{lastpage}} \pubyear{2014}
 \maketitle
 \label{firstpage}

\begin{abstract}
To estimate the peculiar velocity of the Sun with respect to the
Local Standard of Rest (LSR), we used young objects in the Solar
neighborhood with distance measurement errors within 10\%--15\%.
These objects were the nearest Hipparcos stars of spectral classes
O--B2.5, masers with trigonometric parallaxes measured by means of
VLBI, and two samples of the youngest and middle-aged Cepheids.
The most significant component of motion of all these stars is
induced by the spiral density wave. As a result of using all these
samples and taking into account the differential Galactic
rotation, as well as the influence of the spiral density wave, we
obtained the following components of the vector of the peculiar
velocity of the Sun with respect to the LSR:
 $(U_\odot,V_\odot,W_\odot)_\mathrm{LSR}=
 (6.0,10.6,6.5)\pm(0.5,0.8,0.3)\;\mathrm{km\,s}^{-1}$.
We have found that components of the Solar velocity are quite
insensitive to errors of the distance $R_0$ in a broad range of
its values, from $R_0=7.5$~kpc to $R_0=8.5$~kpc, that affect the
Galactic rotation curve parameters. In the same time, the Solar
velocity components
 $(U_\odot)_\mathrm{LSR}$ and
 $(V_\odot)_\mathrm{LSR}$ are very sensitive to the Solar radial phase
$\chi_\odot$ in the spiral density wave.
\end{abstract}


 \section*{Introduction}
The peculiar velocity of the Sun with respect to the LSR
$(U_\odot,V_\odot,W_\odot)_\mathrm{LSR}$ plays an important role
in analysis of the kinematics of stars in the Galaxy. To properly
analyze Galactic orbits, this motion should be removed from the
observed velocities of stars, since it characterizes only the
Solar orbit~--- namely, its deviation from the purely circular
orbit. In particular, to build a Galactic orbit of the Sun, it is
desirable to know the components
$(U_\odot,V_\odot,W_\odot)_\mathrm{LSR}$.

There are several ways to determine the peculiar velocity of the
Sun with respect to the LSR. One of them is based on using the
Str$\ddot{o}$mberg relation. The method consists in finding such
values $(U_\odot,V_\odot,W_\odot)_\mathrm{LSR}$ that correspond to
zero stellar velocity
dispersions~\citep{db98,bob07,ab09,Cockunoglu11,Golubov13}. This
method was addressed, for example, in the work by
\cite{Schonrich10}, where the gradient of metallicity of stars in
the Galactic disk was taken into account, and the velocity
obtained is
 $(U_\odot,V_\odot,W_\odot)_\mathrm{LSR}=
 (11.1,12.2,7.3)\pm(0.7,0.5,0.4)\;\mathrm{km\,s}^{-1}$.

Another method implies a search for such
$(U_\odot,V_\odot,W_\odot)_\mathrm{LSR}$ that lead to minimal
stellar eccentricities. Using this approach, \cite{fa09} suggested
that this velocity is
 $(U_\odot,V_\odot,W_\odot)_\mathrm{LSR}=
 (7.5,13.5,6.8)\pm(1.0,0.3,0.1)\;\mathrm{km\,s}^{-1}$.
To obtain this, 20000 local stars from the New Hipparcos
Reduction~\citep{leuwen07} with known line-of-site velocities were
used. Using the improved database of proper motions and
line-of-site velocities of Hipparcos stars, XHIP, the same authors
found the following new values:
 $(U_\odot,V_\odot,W_\odot)_\mathrm{LSR}=
 (14.2,14.5,7.1)\pm(1.0,1.0,0.1)\;\mathrm{km\,s}^{-1}$
\citep{francis12}; a great effort was made to get
rid of the influence of inhomogeneous distribution of velocities
of stars caused by kinematics of stellar groups and streams.

Another method is based on transferring stellar velocities towards
their origin. Following this approach, \cite{Koval09} derived the
following values of
 $(U_\odot,V_\odot,W_\odot)_\mathrm{LSR}=
 (5.1,7.9,7.7)\pm(0.4,0.5,0.2)\;\mathrm{km\,s}^{-1}.$

Experience in using the Str$\ddot{o}$mberg relation
\citep{db98,bob07,ab09} showed that the youngest stars
significantly deviate from a linear dependence when analyzing
$(V_\odot)_\mathrm{LSR}$. This occurs when the dispersions
$\sigma<17$~km\,s$^{-1} $\citep{db98}. Therefore, the youngest
stars are not normally used in this method. Cepheids and other
youngest objects fall into this area, which allows us to include
them in our analysis. This behavior of velocities of the youngest
stars is primarily connected with the effect of the Galactic
spiral density wave~\citep{LinShu64}.
For instance, the analysis of kinematics of 185 Galactic Cepheids
by~\cite{bb12Ceph} demonstrated that perturbation velocities
inferred by the spiral density wave can be determined with high
confidence.

The purpose of this paper is to estimate the velocity
$(U_\odot,V_\odot,W_\odot)_\mathrm{LSR}$ using spatial velocities
of the youngest objects in the Solar neighborhood that have
parallax errors not larger than 10\%--15\%. For these stars, we
consider not only the impact of the differential rotation of the
Galaxy, but also the influence of the spiral density wave.

\section{Method}\label{method}
Assuming that the angular rotation velocity of the Galaxy
($\Omega$) depends only on the distance  $R$ from the axis of
rotation, $\Omega=\Omega(R)$, the apparent velocity ${\bf V}(r)$
of a star at heliocentric radius ${\bf r}$ can be described in
vectorial notation by the following relation~(Eq.~2.86
in~\cite{ogor65})
 \begin{equation}
 {\bf V}(r)=-{\bf V}_\odot +
 {\bf V_\theta}(R)-{\bf V_\theta}(R_0) +
 {\bf V'},
 \label{Bottlinger-01}\end{equation}
where ${\bf V}_\odot(U_\odot,V_\odot,W_\odot)$ is the mean stellar
sample velocity due to the peculiar Solar motion with respect to
the LSR (hence its negative sign), the velocity $U$ is directed
towards the Galactic center, $V$ is in the direction of Galactic
rotation, $W$ is directed to the north Galactic pole; $R_0$ is the
Galactocentric distance of the Sun; $R$ is the distance of an
object from the Galactic rotation axis;
 ${\bf V_\theta}(R)$ is the circular velocity of the star with respect to the center
 of the Galaxy,
 ${\bf V_\theta}(R_0)$ is the circular velocity of the Sun, while
 ${\bf V'}$ are residual stellar velocities.

 It is necessary to note that Eq.~(\ref{Bottlinger-01}) is widely used by different
 authors for the Galaxy kinematics analysis (for example Eq.~(6)
 in~\cite{Mendez2000} or Eq.~(4) in~\cite{Vallenari06}).

From the above relation~(\ref{Bottlinger-01}), one can write down
three equations in components ($V_r,V_l,V_b$), the so-called
\textit{Bottlinger's equations}~(Eq.~6.27 in \cite{Trumpler53}):
 \begin{equation}
 \begin{array}{lllll}
 V_r= (\Omega-\Omega_0)R_0\sin l \cos b,\\
 V_l= (\Omega-\Omega_0)R_0\cos l-\Omega r\cos b,\\
 V_b=-(\Omega-\Omega_0)R_0\sin l \sin b.
 \label{Bottl-0234}
 \end{array}
 \end{equation}
These are exact formulas, and the signs of $\Omega$ follow
Galactic rotation. The first of these equations was initially
deduced by~\cite{Bottl1931}, while the second one, even earlier,
by~\cite{Pilowski1931}, and the one for $V_b,$~---
by~\cite{ogor48}. After expanding $\Omega$ into Taylor series
against the small parameter $R-R_0$, then expanding the difference
$R-R_0$, where the distance $R$ is
 \begin{equation}
 R^2=r^2\cos^2 b-2R_0 r\cos b\cos l+R^2_0,
 \label{RR}
 \end{equation}
and then substituting the result into Eq.~(\ref{Bottl-0234}), one
gets the equations of the Oort--Lindblad model (Eq.~6.34 in
\cite{Trumpler53}).

Our approach departs from the above in that the distances $r$ are
known quite well. In this case, there is no need to expand $R-R_0$
into series, since the distance $R$ is calculated from
Eq.~(\ref{RR}).

Furthermore, our approach implies an extra assumption that the observed stellar
velocities include perturbations due to the spiral density wave ${\bf V}_{sp}(V_R,\Delta
V_\theta)$, with a linear dependence on both ${\bf V}_{sp}$ and ${\bf V}_\odot$. This
allows us to write
 \begin{equation}
 -{\bf V}_\odot=-{{\bf V}_\odot}_\mathrm{LSR}+{\bf V}_{sp}.
 \label{LSR}\end{equation}
Perturbations from the spiral density wave have a direct influence
on the peculiar Solar velocity ${{\bf V}_\odot}_\mathrm{LSR}$,
which appears to be first pointed out by~\cite{Creze1973}~--- see
the coefficients $a_1$ and $a_2$ in Eq.~(22) of their paper. Then
the relation~(\ref{Bottlinger-01}) takes the following form:
 \begin{equation}
 {\bf V}(r)=-{{\bf V}_\odot}_\mathrm{LSR}+{\bf V}_{sp} +
 {\bf V_\theta}(R)-{\bf V_\theta}(R_0) +
 {\bf V'},
 \label{Bottlinger-02}\end{equation}
which, considering the expansion of the angular velocity of
Galactic rotation $\Omega$ into series up to the second order of
$r/R_0$ reads
\begin{equation}
 \begin{array}{lll}
 V_r=-U_\odot\cos b\cos l-V_\odot\cos b\sin l-W_\odot\sin b\\
 +R_0(R-R_0)\sin l\cos b \Omega^\prime_0\\
 +0.5R_0 (R-R_0)^2 \sin l\cos b \Omega^{\prime\prime}_0\\
 +\Delta V_{\theta}\sin(l+\theta)\cos b-V_R \cos(l+\theta)\cos b,
 \label{EQ-1}
 \end{array}
 \end{equation}
 \begin{equation}
 \begin{array}{lll}
 V_l= U_\odot\sin l-V_\odot\cos l\\
  +(R-R_0)(R_0\cos l-r\cos b) \Omega^\prime_0\\
  +(R-R_0)^2 (R_0\cos l - r\cos b)0.5\Omega^{\prime\prime}_0- r \Omega_0 \cos b\\
  +\Delta V_{\theta} \cos(l+\theta)+V_R\sin(l+\theta),
 \label{EQ-2}
 \end{array}
 \end{equation}
 \begin{equation}
 \begin{array}{lll}
 V_b=U_\odot\cos l \sin b + V_\odot\sin l \sin b-W_\odot\cos b\\
    -R_0(R-R_0)\sin l\sin b\Omega^\prime_0\\
    -0.5R_0(R-R_0)^2\sin l\sin b\Omega^{\prime\prime}_0\\
    -\Delta V_{\theta} \sin (l+\theta)\sin b+V_R \cos(l+\theta)\sin b,
 \label{EQ-3}
 \end{array}
 \end{equation}
where the following designations are used: $V_r$ is the
line-of-sight velocity, $V_l=4.74 r \mu_l\cos b$ and $V_b=4.74 r
\mu_b$ are the proper motion velocity components in the $l$ and
$b$ directions, respectively, with the factor~4.74 being the
quotient of the number of kilometers in an astronomical unit and
the number of seconds in a tropical year; the star's proper motion
components $\mu_l\cos b$ and $\mu_b$ are in mas~yr$^{-1}$, and the
line-of-sight velocity $V_r$ is in km~s$^{-1}$; $\Omega_0$ is the
angular velocity of rotation at the distance $R_0$; parameters
$\Omega^\prime_0$ and $\Omega^{\prime\prime}_0$ are the first and
second derivatives of the angular velocity, respectively. To
account for the influence of the spiral density wave, we used the
simplest kinematic model based on the linear density wave theory
by~\cite{LinShu64}, where the potential perturbation is in the
form of a travelling wave. Then,
 \begin{equation}
      V_R=f_R \cos \chi,
 \label{VR-VR}
 \end{equation}
 \begin{equation}
      \Delta V_{\theta}=f_\theta \sin \chi,
 \label{VR-Vtheta}
 \end{equation}
where $f_R$ and $f_\theta$ are the amplitudes of the radial
(directed toward the Galactic center in the arm) and azimuthal
(directed along the Galactic rotation) velocity perturbations; $i$
is the spiral pitch angle ($i<0$ for winding spirals); $m$ is the
number of arms (we take $m=2$ in this paper); $\theta$ is the
star's position angle measured in the direction of Galactic
rotation: $\tan\theta = y/(R_0-x)$, where $x$ and $y$ are the
Galactic heliocentric rectangular coordinates of the object;
radial phase of the wave $\chi$ is
 \begin{equation}
   \chi=m[\cot (i)\ln (R/R_0)-\theta]+\chi_\odot,
 \label{chi-creze}
 \end{equation}
where $\chi_\odot$ is the radial phase of the Sun in the spiral
density wave; we measure this angle from the center of the
Carina--Sagittarius spiral arm ($R\approx7$~kpc). The parameter
$\lambda$, which is the distance along the Galactocentric radial
direction between adjacent segments of the spiral arms in the
Solar neighborhood (the wavelength of the spiral density wave), is
calculated from the relation
 $2\pi R_0/\lambda = m\cot(i).$

 \begin{figure*}
 \includegraphics[width=2.0\columnwidth]{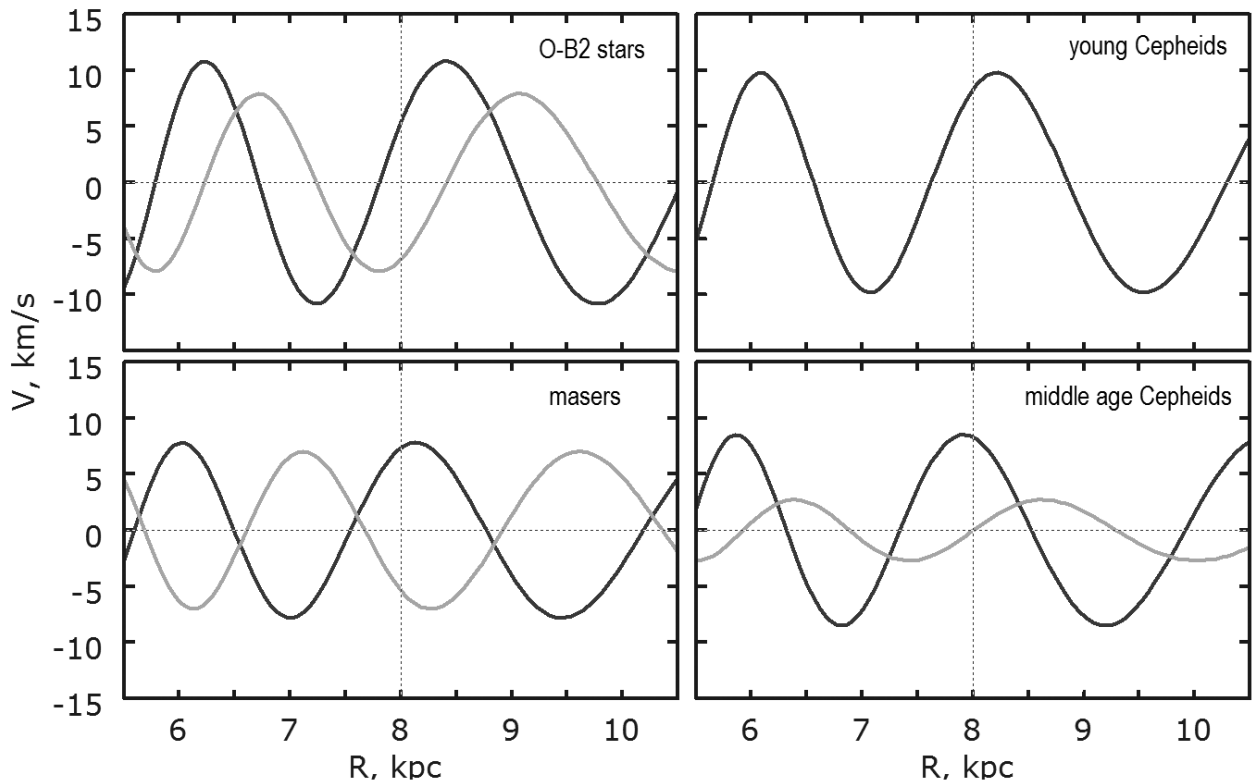}
 \caption{Radial ($V_R$, dark) and tangential ($\Delta V_\theta$, light)
 perturbation velocities versus Galactocentric distances $R$.
 Location of the Sun is indicated by a dotted line.}
  \label{f1}
 \end{figure*}
 \begin{figure}
 \begin{center}
 \includegraphics[width=0.85\columnwidth]{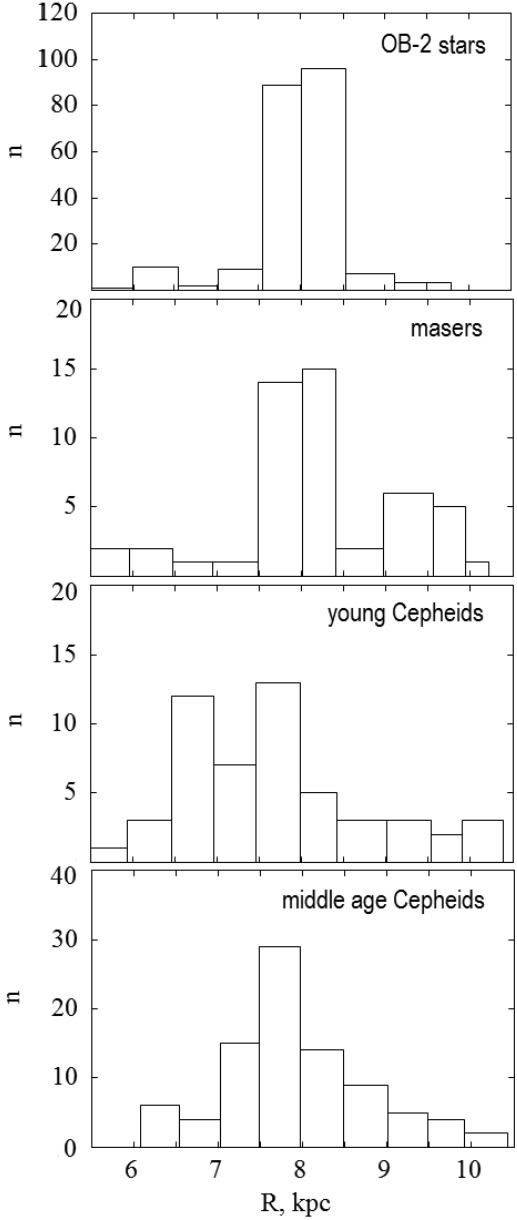}
 \caption{Number of stars versus Galactocentric distances $R$.}
  \label{f2}
  \end{center}
 \end{figure}

We take $R_0=8.0\pm0.4$~kpc, according to analysis of
the most recent determinations of this quantity in the review
by~\cite{FosterR010}.

We use the well-known statistical method~\citep{Trumpler53,ogor65}
to determine the parameters of the residual velocity
(Schwarzschild) ellipsoid. It consists in determining the
symmetric tensor of moments or the tensor of residual stellar
velocity dispersions. When simultaneously using the stellar
line-of-site velocities and proper motions to find the six unknown
components of the dispersion tensor, we have six equations for
each star. The semiaxes of the residual velocity ellipsoid,
denoted by $\sigma_{1,2,3}$, can be determined by analyzing the
eigenvalues of the dispersion tensor.

In the present paper, we assume that parameters of both the
differential Galactic rotation and the spiral density wave are
known from observations of distant stars and solving equations of
the form~(\ref{EQ-1})--(\ref{EQ-3}). In this case the right-hand
parts of the equations contain only components of the Solar
peculiar velocity
\begin{equation}
 \begin{array}{lll}
 V_r-R_0(R-R_0)\sin l\cos b \Omega^\prime_0\\
 -0.5R_0 (R-R_0)^2 \sin l\cos b \Omega^{\prime\prime}_0\\
 -\Delta V_{\theta}\sin(l+\theta)\cos b-V_R \cos(l+\theta)\cos b\\
 =-U_\odot\cos b\cos l-V_\odot\cos b\sin l-W_\odot\sin b,
 \label{EQ-101}
 \end{array}
 \end{equation}
 \begin{equation}
 \begin{array}{lll}
 V_l-(R-R_0)(R_0\cos l-r\cos b) \Omega^\prime_0\\
    -(R-R_0)^2 (R_0\cos l - r\cos b)0.5\Omega^{\prime\prime}_0+r \Omega_0 \cos b\\
    -\Delta V_{\theta} \cos(l+\theta)-V_R\sin(l+\theta)\\
   =U_\odot\sin l-V_\odot\cos l,
 \label{EQ-102}
 \end{array}
 \end{equation}
 \begin{equation}
 \begin{array}{lll}
 V_b+R_0(R-R_0)\sin l\sin b\Omega^\prime_0\\
    +0.5R_0(R-R_0)^2\sin l\sin b\Omega^{\prime\prime}_0\\
    +\Delta V_{\theta} \sin (l+\theta)\sin b+V_R \cos(l+\theta)\sin b\\
  =U_\odot\cos l \sin b + V_\odot\sin l \sin b-W_\odot\cos b.\\
 \label{EQ-103}
 \end{array}
 \end{equation}
The system~(\ref{EQ-101})--(\ref{EQ-103}) can be solved by least-squares
adjustment with respect to three unknowns
$U_\odot,$ $V_\odot$, and $W_\odot$. Another approach (which we follow)
is to calculate components of spatial
velocities $U,V,W$ of stars:
 \begin{equation}
 \begin{array}{lll}
 U=V'_r\cos l\cos b-V'_l\sin l-V'_b\cos l\sin b,\\
 V=V'_r\sin l\cos b+V'_l\cos l-V'_b\sin l\sin b,\\
 W=V'_r\sin b                 +V'_b\cos b,
 \label{EQ-UVW}
 \end{array}
 \end{equation}
where $V'_r,V'_l,V'_b$ are left-hand parts of
Eqs.~(\ref{EQ-101})--(\ref{EQ-103}) which are the observed stellar
velocities free from Galactic rotation and the spiral density
wave. Then
 ${\overline U=-U_\odot},$
 ${\overline V=-V_\odot}$ and
 ${\overline W=-W_\odot}$.

\section{Data}\label{Data}
\subsection{O--B2.5 stars}\label{OB2stars}
The sample of selected 200 massive ($<$10$M_\odot$) stars of
spectral classes O--B2.5 is described in detail in our previous paper
\citep{bb13OB2}. It contains spectral binary O stars with
reliable kinematic characteristics from the $3$~kpc Solar neighborhood.
In addition, the sample contains 124
Hipparcos~\citep{leuwen07} stars of spectral types from B0 to B2.5
whose parallaxes were determined to within 10\% and better
and for which there are line-of-sight velocities in the catalog by
\cite{Gont06}.

In this work we solve the problem of determining the peculiar
velocity of the Sun. This problem can be solved most reliable
using the closest stars to the Sun. Therefore, from the database,
including 200 stars, we have selected 161 stars from the Solar
neighborhood of $0.7$~kpc radius.

Parameters of the Galactic rotation and the spiral density wave
used for reduction of motion of these stars were determined
in~\citep{bb13OB2} using full sample of 200 stars, because for
determination of Galactic parameters it is important to have
distant stars as well:
 $\Omega_0 = 32.4 \pm1.1$~km s$^{-1}$ kpc$^{-1},$
 $\Omega^\prime_0 = -4.33\pm0.19$~km s$^{-1}$ kpc$^{-2},$
 $\Omega^{\prime\prime}_0 = 0.77\pm0.42$~km s$^{-1}$ kpc$^{-3},$
 $f_R=-10.8\pm1.2$~km s$^{-1},$
 $f_\theta= 7.9\pm1.3$~km s$^{-1},$
 $\chi_\odot=-120^\circ\pm4^\circ.$
For all samples in the present work, we use the same value of the
wavelength $\lambda=2.6\pm0.2$~kpc ($i=-6.0\pm0.4^\circ$ for
$m=2$).

\subsection{Masers}\label{masers}
We use coordinates and trigonometric parallaxes of masers measured
by VLBI with errors of less than 10\% in average. These masers are
connected with very young objects (basically proto stars of high
masses, but there are ones with low masses too; a number of
massive super giants are known as well) located in active
star-forming regions.

One of such observational campaigns is the Japanese project VERA
(VLBI Exploration of Radio Astrometry) for observations of water
(H$_2$O) Galactic masers at 22~GHz~\citep{Hirota07} and SiO masers
(which occur very rarely among young objects) at
43~GHz~\citep{Kim08}. Water and methanol (CH$_3$OH) maser
parallaxes are observed in USA (VLBA) at 22~GHz and
12~GHz~\citep{Reid09}. Methanol masers are observed also in the
framework of the European VLBI network~\citep{Rygl10}. Both these
projects are joined together in the BeSSeL
program~\citep{BesseL11}. VLBI observations of radio stars in
continuum at 8.4~GHz~\citep{Dzib11} are carried out with the same
goals.

In the present work, we use only data on the nearby maser sources,
which are located no farther than 1.5~kpc from the Sun. All required
information about 30 such masers is given in the work by~\cite{Xu13},
which is dedicated to the study of the Local arm (the Orion arm).

By applying the reduction algorithm, we use the following
parameters of the Galactic rotation and the spiral density wave
found by~\cite{BobBajk2013Masers}:
 $\Omega_0 = 29.9\pm1.1$~km s$^{-1}$ kpc$^{-1},$
 $\Omega^\prime_0 = -4.27\pm0.20$~km s$^{-1}$ kpc$^{-2},$
 $\Omega^{\prime\prime}_0 = 0.915\pm0.166$~km s$^{-1}$ kpc$^{-3},$
 $f_R=-7.8\pm0.7$~km s$^{-1},$
 $f_\theta= 7.0\pm1.2$~km s$^{-1}$. In this case, the
values of the phase of the Sun in the spiral wave found
independently from radial and tangential perturbations using
Fourier analysis are different:
$(\chi_\odot)_R=-160^\circ\pm15^\circ$ and
$(\chi_\odot)_\theta=-50^\circ\pm15^\circ$, respectively.

Note that line-of-site velocities of masers given in the
literature usually refer to the standard apex of the Sun. So we
fix such line-of-site velocities, making them heliocentric.

 \begin{table*}
 \begin{center}
 \caption{Components of the peculiar velocity of the Sun with respect to the LSR,
 calculated considering the differential Galactic rotation only.}
  \label{t1}
  {\small
   \begin{tabular}{|l|c|c|c|c|l|c|c|c|}      \hline
                    Stars  &    $U_\odot$ &    $V_\odot$ &   $W_\odot$ & $N_\star$ & distance &   $\sigma_1$ &   $\sigma_2$ &  $\sigma_3$ \\
                           &  km s$^{-1}$ &  km s$^{-1}$ & km s$^{-1}$ &           &  kpc     &  km s$^{-1}$ &  km s$^{-1}$ & km s$^{-1}$ \\\hline
 O--B2.5                   & $10.0\pm1.0$ & $14.7\pm1.3$ & $7.2\pm0.7$ &       161 & $<0.7$   & $ 7.3\pm0.5$ & $ 5.7\pm0.6$ & $3.9\pm0.4$ \\
 masers                    & $11.9\pm2.7$ & $16.2\pm3.4$ & $6.2\pm1.7$ &        26 & $<1.5$   & $ 6.0\pm0.7$ & $ 5.7\pm0.6$ & $3.0\pm1.7$ \\
 Cepheids,     $P\geq9^\mathrm{d}$  & $ 6.5\pm2.3$ & $12.0\pm2.4$ & $6.1\pm2.5$ &        36 & $<2$     & $12.2\pm1.3$ & $10.5\pm0.8$ & $5.9\pm4.8$ \\
 Cepheids, $5^\mathrm{d}\leq P<9^\mathrm{d}$ & $ 7.3\pm2.1$ & $11.1\pm2.0$ & $6.4\pm1.8$ &        74 & $<2$     & $14.4\pm2.2$ & $10.3\pm2.0$ & $9.5\pm3.6$ \\\hline
      \end{tabular}}
     \end{center}
   \end{table*}
 \begin{table*}
 \begin{center}
 \caption{Components of the vector of the peculiar velocity of the Sun with respect
  to the LSR, calculated considering both the differential Galactic rotation and the spiral
density wave.}
 \label{t2}
   {\small
   \begin{tabular}{|l|c|c|c|c|l|c|c|c|}      \hline
                    Stars  &     $U_\odot$ &      $V_\odot$ &     $W_\odot$ & $N_\star$ & distance &   $\sigma_1$ &   $\sigma_2$ &  $\sigma_3$ \\
                           &   km s$^{-1}$ &    km s$^{-1}$ &   km s$^{-1}$ &           &  kpc     &  km s$^{-1}$ &  km s$^{-1}$ & km s$^{-1}$ \\\hline
 O--B2.5                   &   $4.6\pm0.7$ &   $ 8.6\pm0.9$ &   $7.2\pm0.7$ &       161 & $<0.7$   & $ 6.0\pm0.5$ & $ 5.8\pm0.5$ & $3.7\pm0.4$ \\
 masers                    &   $6.0\pm1.6$ &   $11.4\pm2.5$ &   $6.2\pm1.7$ &        26 & $<1.5$   & $ 6.2\pm0.4$ & $ 4.8\pm0.6$ & $3.6\pm1.5$ \\
 Cepheids,     $P\geq9^\mathrm{d}$  &   $6.8\pm2.3$ &   $12.1\pm2.4$ &   $6.1\pm2.5$ &        36 & $<2$     & $11.4\pm1.3$ & $10.0\pm0.9$ & $5.6\pm5.1$ \\
 Cepheids, $5^\mathrm{d}\leq P<9^\mathrm{d}$ &   $6.7\pm2.1$ &   $10.4\pm1.9$ &   $6.4\pm1.8$ &        74 & $<2$     & $14.7\pm2.3$ & $ 9.8\pm1.9$ & $9.3\pm3.7$ \\
                   average & $6.0\pm0.5$ & $10.6\pm0.8$ & $6.5\pm0.3$ &           &          \\\hline
      \end{tabular}}
     \end{center}
   \end{table*}

\subsection{Cepheids}\label{cepheids}
We used the data on classical Cepheids with proper motions mainly from
the Hipparcos catalog and line-of-sight velocities from the various
sources. The data from~\cite{Mishurov97} and \cite{Gont06}, as well
as from the SIMBAD database, served as the main sources of
line-of-sight velocities for the Cepheids. For several long-period
Cepheids, we used their proper motions from the TRC~\citep{TRC}
and UCAC4~\citep{UCAC4} catalogs.

To calculate the Cepheid distances, we use the calibration
from~\cite{Fouqu07},
 \begin{equation}\displaystyle
 \langle M_V\rangle=-1.275-2.678\log P,
 \label{Ceph-01}
 \end{equation}
where the period $P$ is in days. Given $\langle M_V\rangle$,
taking the period-averaged apparent magnitudes $\langle V\rangle$
and extinction $A_V=3.23 E(\langle B\rangle-\langle V\rangle)$
mainly from~\cite{Acharova12} and, for several stars,
from~\cite{FeastW97}, we determine the distance $r$ from the
relation
 \begin{equation}\displaystyle
 r=10^{\displaystyle -0.2(\langle M_V\rangle-\langle V\rangle-5+A_V)}
 \label{Ceph-02}
 \end{equation}
and then assume that the relative error of Cepheid distances
determined by this method is 10\%.

We divided the entire sample into two parts, depending on the
pulsation period, which well reflects the mean Cepheid age ($t$).
We use the calibration from~\cite{Efremov03},
\begin{equation}
 \log t=8.50-0.65\log P,
\label{AGE-EFREM}
 \end{equation}
obtained by analyzing Cepheids in the Large Magellanic Cloud.

Parameters of the Galactic rotation and spiral density wave depend
on the age of the Cepheids. Therefore, for each sample of
Cepheids of the given age, these effects should be addressed
individually. We use the values of the parameters found in the
work by~\cite{bb12Ceph} for three age groups. The youngest
Cepheids with periods of $P\geq9^\mathrm{d}$ are characterized by
the average age of $55$~Myr, middle-aged Cepheids with periods of
$5^\mathrm{d}\leq P<9^\mathrm{d}$ have the average age of
$95$~Myr, while the oldest Cepheids with periods of
$P<5^\mathrm{d}$ have that of $135$~Myr. In the present work, a
sample of old Cepheids is not used because there are very few of
them in the Solar neighborhood, and their kinematic parameters are
not very reliable.

According to~\cite{bb12Ceph}, for the youngest Cepheids with periods of
$P\geq9^\mathrm{d}$ $\Omega_0 = 26.1\pm0.9$~km s$^{-1}$ kpc$^{-1}$,
 $\Omega^\prime_0 = -3.95\pm0.13$~km s$^{-1}$ kpc$^{-2}$,
 $\Omega^{\prime\prime}_0 = 0.79\pm0.10$~km s$^{-1}$ kpc$^{-3}$,
 $f_R=-9.8\pm1.3$~km s$^{-1}$,
 $\chi_\odot=-148^\circ\pm14^\circ$,
and the value of velocity perturbations in the tangential
direction $f_\theta$ is assumed to be zero.

For middle-aged Cepheids with $(5^\mathrm{d}\leq P<9^\mathrm{d})$ $\Omega_0 =
30.4\pm1.0$~km s$^{-1}$ kpc$^{-1}$,
 $\Omega^\prime_0 = -4.34\pm0.13$~km s$^{-1}$ kpc$^{-2}$,
 $\Omega^{\prime\prime}_0 = 0.69\pm0.14$~km s$^{-1}$ kpc$^{-3}$,
 $f_R=-8.5\pm1.1$~km s$^{-1}$,
 $f_\theta= 2.7\pm1.1$~km s$^{-1}$. The values for the phase of the
Sun in the spiral wave found separately from radial and tangential
perturbations by periodogram analysis based on Fourier
transform slightly differ: $(\chi_\odot)_R=-193^\circ\pm9^\circ$
and $(\chi_\odot)_\theta=-180^\circ\pm9^\circ.$

In Fig.~\ref{f2}, we show the histograms of stellar distribution
in the samples under analysis versus Galactocentric distances $R$.
To make these plots, we took stars in the $r<2.5$~kpc vicinity.
One can see that the distributions of OB-2.5 stars and masers are
similar. This is due to the fact that nearby masers belong to the
Local arm, while the OB-2.5 star sample is essentially the Gould
belt (since we chose the stars with $\sigma_\pi/\pi<10\%$) that is
a part of the Local arm. Cepheids of different ages manifest the
well-known effect~-- a gradient of the age across the spiral
arm~\citep{PavlSuch78}, which suggests that the spiral arm goes
towards the Galactic center.

 \section{Results and Discussion}\label{Results}
 \subsection{The fixed value of $R_0$}\label{Approach1}
Here we describe the results obtained at fixed value of $R_0=8$
kpc, assuming the parameters of differential Galactic rotation and
the spiral density wave calculated earlier independently for each
stellar sample.

In Figure~\ref{f1}, there are radial ($V_R$) and tangential
($\Delta V_\theta$) velocities of perturbations vs Galactocentric
distance $R$, induced by the spiral density wave. These velocities
are calculated according to the formulas (\ref{VR-VR}),
(\ref{VR-Vtheta}), and (\ref{chi-creze}) assuming
$\theta=0^\circ$, and the amplitudes of perturbations $f_R$ and
$f_\theta$ defined in the data description (Section~\ref{Data}).
As it can be seen from this figure, at $R=R_0$, the perturbations
achieve about $5$~km~s$^{-1}$ in the radial direction. In the
tangential direction, the same value is achieved for two samples:
of youngest O--B2 stars and of masers. In the case of young
Cepheids, perturbations in the tangential direction are not
significant. In the case of middle-aged Cepheids, perturbations in
the tangential direction at $R=R_0$ are close to zero. Note that a
very small Solar neighborhood ($R\rightarrow R_0$) is crucial to
determine the velocity $(U_\odot,V_\odot,W_\odot)_\mathrm{LSR}$.

In Table~\ref{t1}, the components of the peculiar velocity of the
Sun with respect to the LSR
$(U_\odot,V_\odot,W_\odot)_\mathrm{LSR}$ are given. They were
obtained only taking into account the influence of the
differential Galactic rotation. Components of this vector, given
in Table~\ref{t2}, were calculated considering both the effects of
the differential Galactic rotation and of the spiral density wave.
In the last three columns of Tables~\ref{t1}--\ref{t2},  the main
axes of the ellipsoid of residual velocities
$\sigma_1,\sigma_2,\sigma_3$ are given.
We should note that, after considering the perturbations from the
spiral density wave, the values of residual velocity dispersions
$\sigma_1,\sigma_2,\sigma_3$ have slightly different
distributions, which leads to a changed orientation of the
residual velocity ellipsoid. However, a detailed analysis of this
problem is out of the scope of the present study and will be
conducted elsewhere.

As it is seen from Tables~\ref{t1} and \ref{t2}, considering the
effect of the spiral density wave for O--B2.5 stars and for masers
leads to a considerable variation of the components $\Delta
U_\odot$ and $\Delta V_\odot$ by $\approx$6~km$^{-1}$. In
addition, this gives smaller errors of the velocity
$(U_\odot,V_\odot,W_\odot)_\mathrm{LSR}$, which is especially
noticeable for masers.

The velocity $(V_\odot)_\mathrm{LSR}$ (Table~\ref{t1}) found from
the data on masers differs from
$(V_\odot)_\mathrm{LSR}=12.2\;\mathrm{km\,s}^{-1}$\citep{Schonrich10}
by $\approx$4~km\,s$^{-1}$, which is in accordance with the result
of analysis of masers in the Local arm~\citep{Xu13}.

The following average values of the parameters
$(U_\odot,V_\odot,W_\odot)_\mathrm{LSR}$ found in the present work
are, essentially, more accurate than the estimate
 $(U_\odot,V_\odot,W_\odot)_\mathrm{LSR}=
 (5.5,11.0,8.5)\pm(2.2,1.7,1.2)\;\mathrm{km\,s}^{-1}$
obtained from 28 masers by~\cite{bb10} considering the influence
of the spiral density wave. The average value of
$(V_\odot)_\mathrm{LSR}$ (Table~\ref{t2}) is in a good agreement
with the result by~\cite{Schonrich10}. There is a discrepancy in
the $(U_\odot)_\mathrm{LSR}$ component with~\cite{Schonrich10},
and especially with~\cite{francis12}.

Note that the revised Str$\ddot{o}$mberg relation applied to the
experimental RAVE data gives an absolutely different velocity
$(V_\odot)_\mathrm{LSR}\approx3\;\mathrm{km\,s}^{-1}$~\citep{Golubov13}.
 Using another approach to analysis of RAVE data \cite{Pasetto12}
 obtained the following velocities: $(U_\odot,V_\odot)_\mathrm{LSR}=
 (9.87,8.01)\pm(0.37,0.29)\;\mathrm{km\,s}^{-1}$.

Thus different methods give different results, and a final
agreement on the values of the velocity
$(U_\odot,V_\odot,W_\odot)_\mathrm{LSR}$ is not achieved till now.
We consider our estimates most reliable as they are based on the
youngest stars characterized by a small velocity dispersion and by
small Galactic orbit eccentricities as well.

\begin{figure}
 \begin{center}
 \includegraphics[width=0.85\columnwidth]{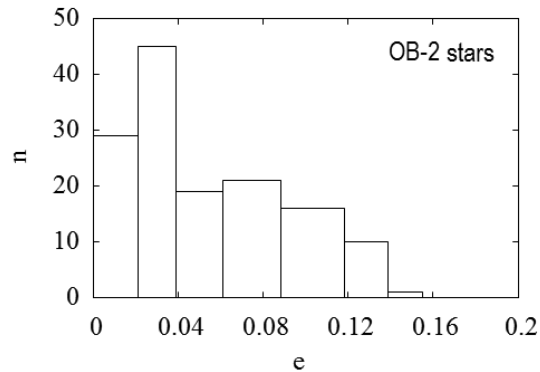}
 \caption{Number of stars versus eccentricity $e$.}
  \label{f3}
  \end{center}
 \end{figure}

In Fig.~\ref{f3}, there is a histogram of Galactic orbital
eccentricities for the OB-2.5 star sample. One can see that
eccentricities of the stars considered are indeed small. Note that
we have excluded escaping stars when making this
sample~\citep{bb13OB2}.

 \begin{table*}
 \begin{center}
 \caption{Components of the vector of the peculiar velocity of the Sun with respect
  to the LSR, calculated considering both the differential Galactic rotation and the spiral
  density wave for three values of $R_0=7.5,8.0,8.5$ kpc. }
 \label{t3}
   {\small
   \begin{tabular}{|l|c|c|c|||c|l|c|c|c|c|c|c|c|}      \hline
 $R_0$                             &             &    $7.5$~kpc &             &&             &    $8.0$~kpc &             &&             &   $8.5$~kpc &             \\\hline
                            Stars  &   $U_\odot$ &    $V_\odot$ &   $W_\odot$ &&   $U_\odot$ &    $V_\odot$ &   $W_\odot$ &&   $U_\odot$ &   $V_\odot$ &   $W_\odot$ \\
                                   & km s$^{-1}$ &  km s$^{-1}$ & km s$^{-1}$ && km s$^{-1}$ &  km s$^{-1}$ & km s$^{-1}$ && km s$^{-1}$ & km s$^{-1}$ & km s$^{-1}$ \\\hline
 O--B2.5                           & $4.9\pm0.5$ & $ 8.6\pm0.5$ & $7.2\pm0.3$ && $4.8\pm0.5$ & $ 8.6\pm0.5$ & $7.2\pm0.3$ && $4.8\pm0.5$ & $ 8.5\pm0.5$ & $7.2\pm0.3$ \\
 masers                            & $5.5\pm1.5$ & $11.3\pm2.5$ & $6.2\pm1.7$ && $6.0\pm1.6$ & $11.4\pm2.5$ & $6.2\pm1.7$ && $5.4\pm1.5$ & $11.5\pm2.5$ & $6.2\pm1.7$ \\
 Ceph.,     $P\geq9^\mathrm{d}$    & $6.1\pm2.3$ & $10.5\pm2.2$ & $6.1\pm2.5$ && $6.6\pm2.5$ & $10.6\pm2.2$ & $6.1\pm2.5$ && $7.0\pm2.6$ & $10.7\pm2.2$ & $6.1\pm2.5$ \\
 $5^\mathrm{d}\leq P<9^\mathrm{d}$ & $7.0\pm2.1$ & $10.8\pm1.9$ & $6.4\pm1.8$ && $6.7\pm2.1$ & $10.8\pm2.0$ & $6.4\pm1.8$ && $6.5\pm2.1$ & $10.9\pm2.0$ & $6.4\pm1.8$ \\\hline
      \end{tabular}}
     \end{center}
   \end{table*}

 \begin{table*}
 \begin{center}
 \caption{Components of the vector of the peculiar velocity of the Sun with respect
  to the LSR, calculated considering both the differential Galactic rotation and the spiral
  density wave for the different values of $\chi_\odot.$
 }
 \label{t4}
   {\small
   \begin{tabular}{|l|c|c|c|||c|l|c|c|c|c|c|c|c||c|c|}      \hline
 $\chi_\odot$ &\multicolumn{2}{|c|} {$-80^\circ$}&\multicolumn{2}{|c|} {$-110^\circ$}&\multicolumn{2}{|c|} {$-120^\circ$}&\multicolumn{2}{|c|} {$-130^\circ$}&\multicolumn{2}{|c|} {$-160^\circ$} \\\hline
              &          $U_\odot$ &    $V_\odot$&            $U_\odot$&    $V_\odot$&            $U_\odot$&    $V_\odot$&            $U_\odot$&    $V_\odot$&            $U_\odot$&    $V_\odot$ \\
              &        km s$^{-1}$ &  km s$^{-1}$&          km s$^{-1}$&  km s$^{-1}$&          km s$^{-1}$&  km s$^{-1}$&          km s$^{-1}$&  km s$^{-1}$&          km s$^{-1}$&  km s$^{-1}$ \\\hline
 O--B2.5      &       $11.4\pm0.5$ &  $7.6\pm0.6$&          $6.3\pm0.5$& $ 8.0\pm0.6$&          $4.8\pm0.5$& $ 8.6\pm0.5$&          $3.5\pm0.5$& $ 9.3\pm0.3$&          $0.8\pm0.6$& $12.5\pm0.5$ \\\hline
      \end{tabular}}
     \end{center}
   \end{table*}

 \subsection{Errors of Galactic Rotation Parameters}
Here we describe the results obtained for three particular values
of $R_0=7.5,8.0,8.5$~kpc using the corresponding differential
Galactic rotation parameters. That is, we now use one and the same
Galactic rotation curve to analyze each of the stellar samples.
Amplitudes of perturbation velocities of the spiral density wave
$f_R$ and $f_\theta$, as well as the values of the Solar phase
$\chi_\odot$ in the spiral wave, are chosen as above in
Section~\ref{Approach1}.

For this purpose, we took a sample of masers (55 masers,
$\sigma_\pi/\pi<10\%,$ $r<3.5$~kpc) from~\cite{BobBajk2013Masers} and,
taking three fixed values of $R_0$, found the following parameters of the
Galactic rotation curve:
\begin{equation}
 \begin{array}{lll}
                     R_0= 7.5~\hbox {kpc,}\\
                \Omega_0= 30.0\pm1.1~\hbox {km s$^{-1}$ kpc$^{-1},$}\\
         \Omega^\prime_0=-4.61\pm0.21~\hbox {km s$^{-1}$ kpc$^{-2},$}\\
 \Omega^{\prime\prime}_0=1.081\pm0.180~\hbox {km s$^{-1}$ kpc$^{-3},$}
  \label{R0-7.5}
 \end{array}
\end{equation}
\begin{equation}
 \begin{array}{lll}
                     R_0= 8.0~\hbox {kpc,}\\
                \Omega_0= 29.9\pm1.1~\hbox {km s$^{-1}$ kpc$^{-1},$}\\
         \Omega^\prime_0=-4.27\pm0.20~\hbox {km s$^{-1}$ kpc$^{-2},$}\\
 \Omega^{\prime\prime}_0=0.915\pm0.166~\hbox {km s$^{-1}$ kpc$^{-3},$}
  \label{R0-8}
 \end{array}
\end{equation}
\begin{equation}
 \begin{array}{lll}
                     R_0= 8.5~\hbox {kpc,}\\
                \Omega_0= 29.8\pm1.1~\hbox {km s$^{-1}$ kpc$^{-1},$}\\
         \Omega^\prime_0=-3.98\pm0.18~\hbox {km s$^{-1}$ kpc$^{-2},$}\\
 \Omega^{\prime\prime}_0=0.783\pm0.154~\hbox {km s$^{-1}$ kpc$^{-3}.$}
  \label{R0-8.5}
 \end{array}
\end{equation}
Parameters~(\ref{R0-8}) are the same as those used previously for the maser
sample in Section~\ref{Approach1}.

The results are summarized in Table~\ref{t3}, where one can see that there is
no considerable departure from the previous results (Table~\ref{t2}). The only
noticeable difference is for the sample of young Cepheids, $\Delta V_\odot\approx2$~km
s$^{-1}$, which is due to the difference in Galactic rotation rate $\Delta \Omega_0\approx2$~km s$^{-1}$.
As early as in the paper of~\cite{FeastW97} it was already noted that the youngest
Cepheids, for some unknown reason, rotate slightly slower than the older ones.
In this sense, we consider the approach taken in the previous paragraph as more
adequate to the goal of the present study: it is better to apply individual rotation
curves to each stellar sample.

 \subsection{Errors of the Spiral Wave Parameters}
Here we describe our results for several model values of the Solar
phase $\chi_\odot$ in the spiral density wave for a sample of
O--B2.5 stars (161 stars, $r<0.7$~kpc). We used the Galactic
rotation curve parameters~(\ref{R0-8}).

The results are reflected in Table~\ref{t4} whence one can see
that the Solar velocity components $U_\odot$ and $V_\odot$ are
very sensitive to the above parameter ($W_\odot$ velocities are
not shown in the Table as they are practically not affected by the
density wave).

It is easy to understand these results by analyzing the corresponding panel of
Fig.~\ref{f1} and Table~\ref{t1}. For instance, for $\chi_\odot=-160^\circ$, the
radial perturbation curve ($V_R$) is near its maximum, so the influence to the
$U_\odot$ component is most prominent. On the contrary, the tangential
perturbation curve ($\Delta V_\theta$) is about zero, so there is no effect on
the $V_\odot$ component.

We must note that, in our previous paper~\citep{bb13OB2}, the
uncertainty of $R_0$ was not taken into account when determining
the Solar phase in the spiral density wave
$\chi_\odot=-120\pm4^\circ$. We have redone Monte Carlo simulation
and obtained the following results:
\begin{enumerate}
 \item If we consider only the error $\sigma_{R_0}=0.4$~kpc, its effect on the
uncertainty of the Solar phase in the spiral density wave becomes
very small: $\sigma_{\chi_\odot}=0.2^\circ$. The explanation for
this is that when you change $R_0,$ the length of a wave stretches
like a rubber band, but the phase of the Sun in the spiral wave
practically does not change.

\item If we consider the errors of all observed parameters of
stars~-- parallaxes, proper motions, line-of-site velocities~--
along with the uncertainty $\sigma_{R_0}$, then the Solar phase in
the spiral density wave becomes $\chi_\odot=-120\pm6^\circ$.
\end{enumerate}
Based on the data from Table~\ref{t4}, we may conclude that, in
the range of phase values from $-110^\circ$ to $-130^\circ$ (which
is even above the $1\sigma$ level), the Solar velocities in
question, found from O--B2.5 stars, are in the
$U_\odot=6-4$~km~s$^{-1}$ and $V_\odot=8-9$~km~s$^{-1}$ range.

\section*{Conclusions}
For evaluation of the peculiar velocity of the Sun with respect to
the Local Standard of Rest, we used young objects from the Solar
neighborhood with distance errors of not larger than 10\%--15\%.
These are the nearest Hipparcos stars of spectral classes O--B2.5,
masers with trigonometric parallaxes measured by means of VLBI,
and two samples of the youngest and middle-aged Cepheids. The
whole sample consists of 297~stars. A significant fraction of
motion of these stars is caused by the Galactic spiral density
wave, because the amplitudes of perturbations in radial ($f_R$)
and tangential ($f_\theta$) directions reach $\approx$10~km
s$^{-1}.$

For each sample of stars, the impact of differential Galactic
rotation and of the Galactic spiral density wave was taken into
account. It was shown that, for the youngest objects~-- namely,
stars of spectral classes O--B2.5 and masers~-- considering the
effect of the spiral density wave leads to a change in the values
of the components of the peculiar velocity of the Sun with respect
to the LSR $\Delta U_\odot$ and $\Delta
V_\odot$ by $\approx$6~km $^{-1}$. Cepheids are less sensitive to
the influence of the spiral density wave.

Average values of the peculiar velocity of the Sun with respect to the
LSR are calculated according to the results of
analysis of four samples of stars; they have the following values:
 $(U_\odot,V_\odot,W_\odot)_\mathrm{LSR}=
 (6.0,10.6,6.5)\pm(0.5,0.8,0.3)$~km~s$^{-1}.$

We have found that components of the Solar velocity are quite
insensitive to errors of the distance $R_0$ in a broad range of
its values, from $R_0=7.5$~kpc to $R_0=8.5$~kpc, that affect the
Galactic rotation curve parameters. In the same time, the Solar
velocity components are very sensitive to the Solar phase
$\chi_\odot$ in the spiral density wave.

In this work we fulfilled data analysis for three different
solar-position/phase values. But it is worth of mentioning an
alternative method based on Bayesian methodology, which has been
used in work by~\cite{McMillan10} for analysis of data on masers.
This method is of interest for us to consider in the future.

\section*{Acknowledgments}
The authors are thankful to the anonymous referee for critical
remarks which promoted improving the paper. The authors are
grateful to L.~P. Ossipkov for a useful discussion. This work was
supported by the ``Nonstationary Phenomena in Objects of the
Universe'' Program of the Presidium of the Russian Academy of
Sciences and the ``Multiwavelength Astrophysical Research'' grant
no. NSh--16245.2012.2 from the President of the Russian
Federation. The authors would like to thank Vladimir Kouprianov
for his assistance in preparing the text of the manuscript.

\end{document}